\documentclass[final,5p,times,twocolumn]{elsarticle}

\usepackage{epsfig}
\newcommand{\ds}{\displaystyle}

\usepackage{amssymb}
\journal{}

\begin{document}

\begin{frontmatter}

%% Title, authors and addresses

%% use the tnoteref command within \title for footnotes;
%% use the tnotetext command for theassociated footnote;
%% use the fnref command within \author or \address for footnotes;
%% use the fntext command for theassociated footnote;
%% use the corref command within \author for corresponding author footnotes;
%% use the cortext command for theassociated footnote;
%% use the ead command for the email address,
%% and the form \ead[url] for the home page:
%% \title{Title\tnoteref{label1}}
%% \tnotetext[label1]{}
%% \author{Name\corref{cor1}\fnref{label2}}
%% \ead{email address}
%% \ead[url]{home page}
%% \fntext[label2]{}
%% \cortext[cor1]{}
%% \address{Address\fnref{label3}}
%% \fntext[label3]{}

\title{Effects of Two Successive Parity-Invariant Point Interactions 
on One-Dimensional Quantum Transmission: 
Resonance Conditions for the Parameter Space}

%% use optional labels to link authors explicitly to addresses:
%% \author[label1,label2]{}
%% \address[label1]{}
%% \address[label2]{}

\author{Kohkichi Konno}
\ead{kohkichi@tomakomai-ct.ac.jp}

\author{Tomoaki Nagasawa}
\ead{nagasawa@tomakomai-ct.ac.jp}

\author{Rohta Takahashi}
\ead{takahashi@tomakomai-ct.ac.jp}

\address{National Institute of Technology, Tomakomai College, 
             443 Nishikioka, Tomakomai 059-1275, Japan}

\begin{abstract}
%% Text of abstract
We consider the scattering of a quantum particle 
by two independent, successive parity-invariant 
point interactions in one dimension. 
The parameter space for the two point interactions
is given by the direct product of two tori, 
which is described by four parameters. 
By investigating the effects of the two point interactions
on the transmission probability of plane wave, 
we obtain the conditions for the parameter space 
under which perfect resonant transmission occur. 
The resonance conditions are found to be described by 
symmetric and anti-symmetric relations between the parameters.
\end{abstract}

\begin{keyword}
%% keywords here, in the form: keyword \sep keyword

one-dimensional quantum systems \sep transmission \sep resonance  

%% PACS codes here, in the form: \PACS code \sep code

\PACS 03.65.-w \sep 03.65.Xp \sep 03.65.Db

%% MSC codes here, in the form: \MSC code \sep code
%% or \MSC[2008] code \sep code (2000 is the default)

\end{keyword}

\end{frontmatter}

%% \linenumbers

%% main text
\section{Introduction}
\label{sec1}

The existence of various non-trivial junction conditions 
for a point interaction in one-dimensional quantum systems
is an intriguing aspect in quantum mechanics.
The property of the junction conditions was fully revealed 
by the mathematical works \cite{rs,seba,aghh,cft} 
and has also been pointed out by a number of research 
\cite{d1,d2,d3,d4,d5,d6,d7,d8,d9,d10,
d11,d12,d13,d14,d15,d16,d17,d18,d19,d20,d21} 
on one-dimensional quantum systems with potential barriers 
made of the Dirac delta function and its (higher) derivatives
(see \cite{lange} for a new approach based on the integral form).
The point interaction in one-dimensional quantum systems
has a relatively large parameter space, in comparison 
with those in higher dimensions. It has been known that 
the parameter space in one dimension is characterized
by $U(2)$, while those in two dimensions and 
three dimensions are characterized by $U(1)$. 
Several authors \cite{b1,b2,b3,b4,b5,b6,n1,n2,n3,b7,b8} reported  
that the interesting characteristics of supersymmetry,
geometric phase, anholonomy, duality, and so on appear
owing to the large parameter space for the junction conditions 
in one dimension.
These previous works placed a special emphasis relatively
on bound states in one-dimensional systems.
Thus, we now consider the scattering of a 
quantum particle by point interactions in one dimension.
The essential properties of the scattering by a
single point interaction were discussed in \cite{cft,b3}. 
Furthermore, it was shown in \cite{hc} that the quantum 
transmission through arbitrarily located $N$ 
point interactions that have scale invariance
exhibits random quantum dynamics.
In this paper, focusing on quantum resonance,
we investigate the occurrence of resonant transmission 
through two independent, successive point interactions.

As for the resonant tunneling, it is remarkable that 
a property inherent in quantum mechanics 
plays a crucial role in this phenomena.
Since the leading work in \cite{bohm}, the basic features 
had been investigated theoretically \cite{iogansen,te} 
and experimentally \cite{cet}. These studies have motivated 
various subsequent works; realistic effects on the resonant tunneling 
were discussed in \cite{ra,yama,ferry,razavy,dmag,yoneta,kpsj}, 
and some different theoretical methods which can deal with 
an arbitrary finite periodic potential were developed 
in \cite{fp1,fp2,fp3,fp4,fp5,fp6}.
Furthermore, the resonant tunneling is still an active area 
of research for the applications to high-frequency oscillators 
in recent years \cite{satsy,fscm,fksa}.
By virtue of recent technology, i.e., nanotechnology, 
the microfabrication down to the atomic scale becomes possible, 
and one-dimensional conductors also become accessible.
However, the effects of the above-mentioned non-trivial 
junction conditions in one dimensional quantum systems 
on resonant transmission have not been fully discussed
in the literature.

The parameter space for two independent, successive 
point interactions in one-dimensional quantum systems
is given by $U(2) \otimes U(2)$. Thus two point 
interactions are characterized by eight parameters.
In this paper, we particularly pay our attention to 
the important subclass for junction conditions which has 
parity invariance and includes typical junction 
conditions, like that for a free particle with no interaction,
that for a delta function potential, and that for 
a epsilon function potential.
When we consider this subclass, the parameter space of 
each point interaction is given by a torus $T^2=S^1 \otimes S^1$,
and thus the parameter space of two independent, 
successive point interactions is reduced to $T^2 \otimes T^2$,
which is described by four parameters.
Nevertheless, even in this reduced parameter space, 
whether resonant transmission occurs or not is 
quite non-trivial. Thus, we investigate the conditions 
for the parameter space under which the resonant transmission 
occur in one-dimensional quantum systems with two 
successive parity-invariant point interactions

This paper is organized as follows.
In Sec.~\ref{sec2}, we review the junction conditions
for a point interaction in one-dimensional quantum systems
and discuss the scattering of plane wave 
by a parity-invariant point interaction.
In Sec.~\ref{sec3}, we deal with quantum transmission
through two different, successive point 
interactions, and investigate the conditions for the 
parameter space under which perfect resonant 
transmission occur. Finally, we give concluding 
remarks in Sec.~\ref{sec4}.

\section{One-dimensional quantum systems with a 
         parity-invariant point interaction}
\label{sec2}

\subsection{The Schr\"odinger equation and junction conditions}

\begin{figure}[tb]
  \includegraphics[width=.45 \textwidth]{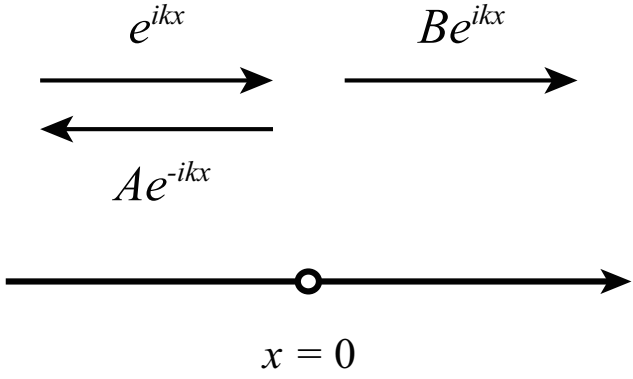}
  \caption{One dimensional space with a point interaction,  
           which is located at $x=0$. The incident wave from the
           left-hand side is scattered by the point of $x=0$.}
  \label{fig1}
\end{figure}

We consider quantum mechanics in one spatial dimension
(say, $x$-axis) with a point interaction 
located at the origin ($x=0$) (see Fig.~\ref{fig1}). 
The wave function $\psi (t, x)$ is governed by 
the Schr\"odinger equation
\begin{eqnarray}
\label{eq:schroedinger}
 i\hbar \frac{\partial}{\partial t} \psi \left( t, x \right)
 = - \frac{\hbar^2}{2m}  \frac{\partial^2}{\partial x^2} 
   \psi \left( t, x \right)
   \quad \left( x \in \mathbb{R} \backslash \{0\} \right),
\end{eqnarray}
where $i$, $\hbar$ and $m$ denote the imaginary unit, 
the Plank constant and the mass of a particle, respectively. 
The probability current is expressed as
\begin{eqnarray}
\label{eq:p-current}
 j (t,x) & = & \frac{\hbar}{2mi} 
  \left\{ \psi^{\ast} (t,x)  \frac{\partial}{\partial x} \psi (t,x) 
   - \psi (t,x) \frac{\partial}{\partial x} \psi^{\ast} (t,x)  \right\},
   \nonumber \\
\end{eqnarray}
where $(^{\ast})$ denotes the complex conjugate.

The junction condition at the point interaction 
is provided by the conservation of the probability current
%%%1
\footnote{This condition is equivalent to 
that derived from the choice of 
a self-adjoint extension of the Hamiltonian.
See the comment below Eq.~(\ref{eq:Phi-c2}).}
%%%
\begin{eqnarray}
\label{eq:j-convervation}
 j (-0)=j(+0),
\end{eqnarray}
where $+0$ and $-0$ denote the limits 
to zero from above and below, respectively,
and the time variable $t$ is abbreviated from now on.
Substituting Eq.~(\ref{eq:p-current}) into Eq.~(\ref{eq:j-convervation}), 
we derive
\begin{eqnarray}
\label{eq:jc0}
 \lefteqn{
  \psi^{\ast} (-0) \psi'(-0)-\psi (-0) {\psi^{\ast}}'(-0)
  } \nonumber \\
  && 
  = \psi^{\ast} (+0) \psi'(+0)-\psi (+0) {\psi^{\ast}}'(+0) ,
\end{eqnarray}
where the prime $(')$ denotes the differentiation with respect to $x$.
When we introduce new vectors as in \cite{cft},
\begin{equation}
 \label{eq:def-Phi}
 \Psi := \left( \begin{array}{c}
	         \psi (+0) \\ \psi (-0)
		\end{array}\right) , \quad 
 \Psi' := \left( \begin{array}{c}
	         \psi' (+0) \\ -\psi' (-0)
		\end{array}\right) ,
\end{equation}
Eq.~(\ref{eq:jc0}) can be expressed as
\begin{equation}
 \label{eq:Phi-c}
 \Psi'^{\dagger} \Psi = \Psi^{\dagger} \Psi' ,
\end{equation}
where $(^{\dagger})$ denotes the transpose of the complex conjugate.
Equation (\ref{eq:Phi-c}) is equivalently expressed as
\begin{equation}
 \label{eq:Phi-c2}
 \left| \Psi - i L_{0} \Psi' \right| = 
  \left| \Psi + i L_{0} \Psi' \right|
\end{equation}
where $L_{0} \left( \in \mathbb{R} \right)$ 
is an arbitrary nonvanishing constant with the
dimension of length. Thus, $\Psi - i L_{0} \Psi'$
is connected to $\Psi + i L_{0} \Psi'$ 
via a unitary transformation.
%%%2
Note that the condition (\ref{eq:Phi-c2}) was 
derived also from the method of a self-adjoint 
extension of the Hamiltonian in \cite{bonneau},
although the notation is slightly different from ours.
%%%
Therefore, we obtain the junction condition \cite{cft}
\begin{equation}
 \label{eq:jc}
 (U-I) \Psi + i L_{0}(U+I) \Psi' = 0 ,
\end{equation}
where $I$ is the $2\times 2$ identity matrix, and
$U$ is a $2\times 2$ unitary matrix, i.e., $U\in U(2)$.

It is sometimes useful to adopt 
the following parametrization for $U$,
\begin{equation}
\label{eq:u-matrix}
 U=e^{i\xi I} e^{i\zeta \sigma_1} e^{i\eta \sigma_2} 
   e^{i\chi \sigma_3} ,
\end{equation}
where $\sigma_{i} \ (i=1, 2, 3)$ denotes the Pauli matrices, 
and $\xi, \zeta, \eta, \chi \left( \in \mathbb{R} \right)$
are parameters.
For example, when we take $\xi=\pi / 2$, 
$\zeta=- \pi / 2$, $\eta =\chi =0$, 
we retrieve a free particle with no interaction, in which
$\psi(-0)=\psi(+0)$, $\psi'(-0)=\psi'(+0)$. When we take
$\xi=\left( \theta +\pi \right)/2$, 
$\zeta=\left( \theta -\pi \right)/2$, $\eta =\chi =0$,
where $\theta$ is a parameter, we can derive 
a potential made of the Dirac delta function $\delta (x)$.

\subsection{Parity-invariant junction conditions}

We restrict our attention to the parity-invariant junction conditions.

We now introduce the parity transformation ${\cal P}$,
which acts on the wave function as
\begin{equation}
 \label{eq:parity-t}
 {\cal P} \psi (x) = \psi (-x) .
\end{equation}
Since ${\cal P}^2 \psi (x) = \psi (x)$, the eigenvalues of ${\cal P}$
take $\pm 1$. We assume the eigenstates to be $\psi_{+}$
and $\psi_{-}$ for the eigenvalues $+1$ and $-1$, respectively,
i.e.,
\begin{equation}
 {\cal P} \psi_{\pm} (x) = \pm \psi_{\pm} (x) .
\end{equation}
The eigenstates $\psi_{\pm}$ are found to be
\begin{equation}
 \label{eq:pi-eigenf}
 \psi_{\pm} (x) = \frac{\psi(x)\pm \psi(-x)}{2}.
\end{equation}
The parity transformations of $\Psi$ and $\Psi'$
are given, respectively, by
\begin{equation}
 \Psi \stackrel{\cal P}{\longrightarrow} \sigma_{1} \Psi,
 \quad \mbox{and} \quad 
 \Psi' \stackrel{\cal P}{\longrightarrow} \sigma_{1} \Psi'.
\end{equation}
We define the projection operators ${\cal P}_{\pm}$
onto the states $\psi_{\pm} $ as 
\begin{equation}
\label{eq:def-ppm}
 {\cal P}_{\pm}:= \frac{I\pm \sigma_{1}}{2},
\end{equation}
so that we have
\begin{equation}
 \label{eq:P_Phi}
 {\cal P}_{+} \Psi = \left( \begin{array}{c}
	         \psi_{+} (+0) \\ \psi_{+} (+0)
		\end{array}\right) , \quad 
 {\cal P}_{-} \Psi = \left( \begin{array}{c}
	         \psi_{-} (+0) \\ -\psi_{-} (+0)
		\end{array}\right) .
\end{equation}
These projection operators satisfy the relations
\begin{equation}
\label{eq:p-rel1}
 \left( {\cal P}_{\pm} \right)^2 = {\cal P}_{\pm}, 
\end{equation}
\begin{equation}
\label{eq:p-rel2}
 {\cal P}_{\pm} {\cal P}_{\mp} =0, 
\end{equation}
\begin{equation}
\label{eq:p-rel3}
 {\cal P}_{+} + {\cal P}_{-} = I .
\end{equation}
The parity transformation of the junction 
condition (\ref{eq:jc}) becomes
\begin{equation}
 \label{eq:jc2}
 (\sigma_{1} U \sigma_{1} -I) \sigma_{1} \Psi 
  + i L_{0}(\sigma_{1} U \sigma_{1} +I) \sigma_{1} \Psi' = 0 ,
\end{equation}
where $\sigma_1$ is multiplied from the left-hand side.
Thus the unitary matrix $U$ is transformed 
under the parity transformation as
\begin{equation}
 U \stackrel{\cal P}{\longrightarrow} \sigma_{1} U \sigma_{1} .
\end{equation}
Therefore, the parity invariance imposes the condition 
%%%3
\footnote{
The authors of \cite{bonneau} derived a boundary 
condition from the method of a self-adjoint extension 
in a system of infinitely deep well potential.
Their condition can be expressed in our notation as
$\sigma_{1} U^{T} \sigma_{1} =U$, where the superscript $T$
denotes the transpose. Thus, their boundary 
condition corresponds to that 
for PT (parity and time-reversal) invariance 
(see also \cite{cft}).
}
%%%
\begin{equation}
 \label{eq:u-condition}
 \sigma_{1} U \sigma_{1} =U 
\end{equation}
on the unitary matrix $U$ for the junction condition.

We can easily show that the unitary matrix $U_{\rm p}$ 
satisfying the parity-invariant condition  (\ref{eq:u-condition}) 
is given by $\eta=\chi=0$ 
for the parametrization of Eq.~(\ref{eq:u-matrix}), i.e.,
\begin{equation}
\label{eq:up-matrix}
 U_{\rm p}=e^{i\xi I} e^{i\zeta \sigma_1} .
\end{equation}
This class of unitary matrices includes 
the junction condition for a free particle with no interaction 
and that for a delta function potential.

Let us derive the parity-invariant junction conditions
for the wave function explicitly.
For our purpose, we rewrite 
$U_{\rm p}$ in Eq.~(\ref{eq:up-matrix})  as
\begin{equation}
\label{eq:up-matrix2}
 U_{\rm p}=e^{i\theta_{+}}{\cal P}_{+} 
           +e^{i\theta_{-}}{\cal P}_{-} ,
\end{equation}
where we define 
\begin{equation}
 \theta_{\pm}:=\xi \pm \zeta .
\end{equation}
These parameters $\theta_{\pm}$ describe a torus
$T^2 =S^1 \otimes S^1$.
Here we have used Eqs.~(\ref{eq:def-ppm}), 
(\ref{eq:p-rel1})--(\ref{eq:p-rel3}) and 
the Baker-Campbell-Hausdorff relation \cite{gr}
\begin{eqnarray}
 e^{X}e^{Y} 
 & = & \exp \Big( X+Y +\frac{1}{2} \left[ X,Y \right] 
  \nonumber \\
  &&
  + \frac{1}{12} \left( \left[ \left[ X,Y \right], Y \right]
   + \left[ X ,\left[ X,Y \right] \right] \right) 
  + \cdots \Big) ,
\end{eqnarray}
where $\left[ X,Y\right]:=XY-YX$. 
 Substituting Eq.~(\ref{eq:up-matrix2})
into Eq.~(\ref{eq:jc}), we derive the junction condition
\begin{eqnarray}
 \label{eq:jc-pi}
 \lefteqn{
  \left( e^{i\theta_{+}}-1 \right) {\cal P}_{+} \Psi 
  + i L_{0} \left( e^{i\theta_{+}}+1 \right) {\cal P}_{+} \Psi'
  } \nonumber \\
  &
  +\left( e^{i\theta_{-}}-1 \right) {\cal P}_{-} \Psi 
  + i L_{0} \left( e^{i\theta_{-}}+1 \right) {\cal P}_{-} \Psi' = 0 .
\end{eqnarray}
Here we have
\begin{equation}
 \label{eq:P_dPhi}
 {\cal P}_{+} \Psi' = \left( \begin{array}{c}
	         \psi'_{+} (+0) \\ \psi'_{+} (+0)
		\end{array}\right) , \quad 
 {\cal P}_{-} \Psi' = \left( \begin{array}{c}
	         \psi'_{-} (+0) \\ -\psi'_{-} (+0)
		\end{array}\right) .
\end{equation}
The junction condition (\ref{eq:jc-pi}) can be divided into two parts;
one is derived by multiplying Eq.~(\ref{eq:jc-pi}) by ${\cal P}_{+}$
from the left-hand side, and the other is derived by multiplying 
Eq.~(\ref{eq:jc-pi}) by ${\cal P}_{-}$ in the same way. 
The resultant equations are
\begin{eqnarray}
 \label{eq:jc-phi1}
 \left( e^{i\theta_{+}}-1 \right) {\cal P}_{+} \Psi 
  + i L_{0} \left( e^{i\theta_{+}}+1 \right) {\cal P}_{+} \Psi' = 0 ,
\end{eqnarray}
\begin{eqnarray}
 \label{eq:jc-phi2}
 \left( e^{i\theta_{-}}-1 \right) {\cal P}_{-} \Psi 
  + i L_{0} \left( e^{i\theta_{-}}+1 \right) {\cal P}_{-} \Psi' = 0 .
\end{eqnarray}
Substituting Eqs.~(\ref{eq:P_Phi}) and (\ref{eq:P_dPhi}) 
into Eqs.~(\ref{eq:jc-phi1}) and (\ref{eq:jc-phi2}), we derive
\begin{eqnarray}
 \label{eq:jc-psi1}
 \psi_{+} (+0) +L^{(+)} \psi'_{+} (+0) =0 ,
\end{eqnarray}
\begin{eqnarray}
 \label{eq:jc-psi2}
 \psi_{-} (+0) +L^{(-)} \psi'_{-} (+0) =0 ,
\end{eqnarray}
where $L^{( \pm )} \left( \in \mathbb{R} \right)$ 
are defined as
\begin{eqnarray}
 L^{( \pm )} := L_{0} \cot \frac{\theta_{\pm}}{2} .
\end{eqnarray}
When we use Eq.~(\ref{eq:pi-eigenf}),  
Eqs.~(\ref{eq:jc-psi1}) and (\ref{eq:jc-psi2}) 
are expressed as
%%%4
\footnote{
These boundary conditions were also obtained in \cite{kurasov}  
from the self-adjoint extension.
Equation (18) in \cite{kurasov} 
under the conditions of $X_{1} =-\frac{2}{L^{(+)}}$, 
$X_{2}=X_{3}=0$, and $X_{4}=2L^{(-)}$  
corresponds to our boundary conditions 
in Eqs. (\ref{eq:jc-psi3}) and (\ref{eq:jc-psi4}).
}
%%%
\begin{eqnarray}
 \label{eq:jc-psi3}
 \left( \psi (+0) +\psi (-0) \right)
 +L^{(+)} \left( \psi' (+0) -\psi' (-0) \right) =0 ,
\end{eqnarray}
\begin{eqnarray}
 \label{eq:jc-psi4}
 \left( \psi (+0) -\psi (-0) \right)
 +L^{(-)} \left( \psi' (+0) +\psi' (-0) \right) =0 .
\end{eqnarray}
Consequently, Eqs.~(\ref{eq:jc-psi3}) and (\ref{eq:jc-psi4}) 
provide the parity-invariant junction conditions
for the wave function.

We provide characteristic examples for the parity-invariant 
junction conditions.
\begin{enumerate}
 \item[(i)] {\it Decoupling boundary conditions 
(Robin boundary conditions).}--- 
When $L^{(+)} = L^{(-)}=L$, the junction conditions
(\ref{eq:jc-psi3}) and (\ref{eq:jc-psi4}) reduce to
\begin{eqnarray}
 \psi (+0) + L \psi' (+0) =0 ,
\end{eqnarray}
\begin{eqnarray}
 \psi (-0) - L \psi' (-0) =0 .
\end{eqnarray}
These leads to $j(+0)=j(-0)=0$. Thus,
the probability current vanishes at $x=0$.
Therefore, the wave function in $x<0$ is completely 
decoupled from that in $x>0$ in this case. 

\item[(ii)] {\it Scale-invariant boundary conditions.}--- 
The scale-invariant feature appears in the following cases:
 \begin{enumerate}
  \item When $\theta_{+}=\theta_{-}=0$, i.e., 
  $L^{(+)} \rightarrow \infty \ (\mbox{or} \ -\infty)$ 
  and $L^{(-)} \rightarrow \infty \ (\mbox{or} \ -\infty)$,
  we derive
  \begin{equation}
   \psi'(+0) = \psi'(-0) =0.
  \end{equation}
  This is the Neumann boundary condition.

  \item When $\theta_{+}=\theta_{-}=\pi$, i.e., 
  $L^{(+)} = L^{(-)} =0$,
  we derive
  \begin{equation}
   \psi(+0) = \psi(-0) =0.
  \end{equation}
  This is the Dirichlet boundary condition.

  \item When $\theta_{+}=0$ and $\theta_{-}=\pi$, i.e., 
  $L^{(+)} \rightarrow \infty \ (\mbox{or} \ -\infty)$ and $L^{(-)} =0$,
  we derive
  \begin{equation}
   \psi(+0) = \psi(-0) ,  \
   \mbox{and} \ \psi'(+0) = \psi'(-0) .
  \end{equation}
  This gives a free particle with no interaction.

  \item When $\theta_{+}=\pi$ and $\theta_{-}=0$, i.e., 
  $L^{(+)} =0$ and
  $L^{(-)} \rightarrow \infty \ (\mbox{or} \ -\infty)$,
  we derive
  \begin{equation}
   \psi(+0) = -\psi(-0) ,  \
   \mbox{and} \ \psi'(+0) = -\psi'(-0) .
  \end{equation}
  This induces the phase inversion at the boundary.
 \end{enumerate}

\item[(iii)] {\it Boundary conditions of the Dirac delta function.}---
When $\theta_{-}=\pi$, i.e., $L^{(-)} =0$,  we derive
  \begin{equation}
   \psi(+0) = \psi(-0) , 
  \end{equation}
and 
  \begin{equation}
   \psi'(+0) -\psi'(-0) = - \frac{2}{L^{(+)}} \psi (+0).
  \end{equation}
This gives a potential by the Dirac delta function.
\end{enumerate}

\subsection{Scattering of plane wave}

We discuss the scattering of plane wave approaching
from the region of $x<0$ by the point interaction
as shown in Fig.~\ref{fig1}. 
%%%5
(See also \cite{boya}, which is an excellent review.)
%%%
We assume the wave function as
\begin{equation}
\label{eq:ef-psi1}
 \psi (x) = \left\{ 
    \begin{array}{ll}
	e^{ikx} + A e^{-ikx}  & (x<0) \\
	Be^{ikx} & (x>0) 	    
    \end{array} \right. ,
\end{equation}
where $k (>0)$ denotes the wave number, and
$A, B \left( \in \mathbb{C} \right)$ are constants 
which are determined by the junction conditions.
When we adopt the junction conditions 
(\ref{eq:jc-psi3}) and (\ref{eq:jc-psi4}) at $x=0$
for the wave function in Eq.~(\ref{eq:ef-psi1}),
we obtain 
\begin{eqnarray}
 A & = &  - \frac{1+ k^2 L^{(+)} L^{(-)}}
    {\left( 1+ ikL^{(+)}\right) \left( 1+ikL^{(-)}\right)} ,\\
 B & = & \frac{ik \left( L^{(+)} - L^{(-)}\right)}
    {\left( 1+ ikL^{(+)}\right) \left( 1+ikL^{(-)}\right)} .
\end{eqnarray}
Note that the same expressions are obtained when 
the plane wave approaches from the region of $x>0$.
This is the natural result from the parity invariance.
The transmission probability $T_{1}$ is 
calculated as
\begin{eqnarray}
\label{eq:t-single}
 T_{1} = \left| B \right|^2 
 = \frac{k^2 \left( L^{(+)} - L^{(-)}\right)^2}
 {\left( 1+ k^2\left( L^{(+)} \right)^2 \right) 
  \left( 1+ k^2\left( L^{(-)} \right)^2 \right)}  .
\end{eqnarray}
It is interesting that $T_{1}$ decreases to zero 
as $k \rightarrow \infty$ in most cases if $L^{(+)}\neq 0$ 
and $L^{(-)}\neq 0$. This fact defies our intuition,
because even a high energy particle could not 
penetrate the potential barrier.
 From the inequality $T_{1}\leq 1$, we also derive
\begin{eqnarray}
\label{eq:ineq-single}
 \left( L^{(+)}  L^{(-)} k^2 +1 \right)^2 \geq 0  .
\end{eqnarray}
Therefore, while the transmission probability $T_{1}$
completely vanishes when $L^{(+)}=L^{(-)}$,
the perfect transmission (i.e., $T_{1}=1$) occurs
when $k=\sqrt{-1/ \left( L^{(+)} L^{(-)} \right)}$
if $L^{(+)} L^{(-)} <0$.

\section{One-dimensional quantum systems with two 
         parity-invariant point interactions}
\label{sec3}

\subsection{Scattering of plane wave by 
     two parity-invariant point interactions}

\begin{figure}[tb]
  \includegraphics[width=.45 \textwidth]{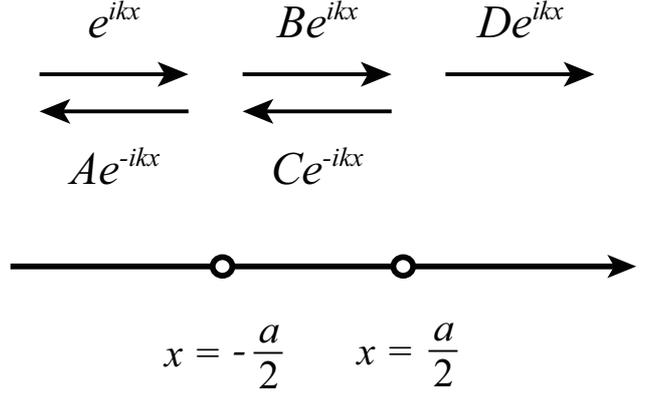}
  \caption{One dimensional space with two point interactions,  
           which are located at $x=-a/2$ and $x=a/2$. 
           The incident wave from the left-hand side 
           is scattered by the points of $x=-a/2$ and $x=a/2$.}
  \label{fig2}
\end{figure}

Let us discuss quantum mechanics in one spatial dimension
with two point interactions, which are located 
at $x=-a/2$ and $x=a/2$ (see Fig.~\ref{fig2}). 
The wave function is assumed to be 
\begin{equation}
\label{eq:ef-psi2}
 \psi (x) = \left\{ 
    \begin{array}{ll}
	e^{ikx} + A e^{-ikx}  & 
          \left( x< - \ds\frac{a}{2} \right) \\
	Be^{ikx} + Ce^{-ikx} & 
          \left( - \ds\frac{a}{2} < x<
	   \ds\frac{a}{2} \right) \\
        De^{ikx} & 
          \left( \ds\frac{a}{2} < x \right) 
    \end{array} \right. ,
\end{equation}
where $A, B , C, D \left( \in \mathbb{C} \right)$ are constants.
In the same way as in Eqs.~(\ref{eq:jc-psi3}) and (\ref{eq:jc-psi4}),
the parity-invariant junction conditions at $x=-a/2$ and $x=a/2$ become, 
respectively,
\begin{eqnarray}
 \label{eq:jc-psi5}
 \lefteqn{
  \left\{ \psi \left( -\ds\frac{a}{2} +0 \right) 
  +\psi \left( -\ds \frac{a}{2} -0 \right) \right\}
  } \nonumber \\
 & 
   +L^{(+)}_{1} \left\{ \psi' \left( -\ds \frac{a}{2} +0 \right) 
   -\psi' \left( -\ds\frac{a}{2} -0 \right) \right\} =0 ,
\end{eqnarray}
\begin{eqnarray}
 \label{eq:jc-psi6}
 \lefteqn{
  \left\{ \psi \left( -\ds\frac{a}{2} +0 \right) 
  -\psi \left( -\ds \frac{a}{2} -0 \right) \right\}
  } \nonumber \\
 & 
   +L^{(-)}_{1} \left\{ \psi' \left( -\ds \frac{a}{2} +0 \right) 
   +\psi' \left( -\ds\frac{a}{2} -0 \right) \right\} =0 ,
\end{eqnarray}
and 
\begin{eqnarray}
 \label{eq:jc-psi7}
 \lefteqn{
   \left\{ \psi \left( \ds\frac{a}{2} +0 \right) 
   +\psi \left( \ds \frac{a}{2} -0 \right) \right\}
   } \nonumber \\
  & 
   +L^{(+)}_{2} \left\{ \psi' \left( \ds \frac{a}{2} +0 \right) 
   -\psi' \left( \ds\frac{a}{2} -0 \right) \right\} =0 ,
\end{eqnarray}
\begin{eqnarray}
 \label{eq:jc-psi8}
\lefteqn{
  \left\{ \psi \left( \ds\frac{a}{2} +0 \right) 
  -\psi \left( \ds \frac{a}{2} -0 \right) \right\}
  } \nonumber \\
 & 
   +L^{(-)}_{2} \left\{ \psi' \left( \ds \frac{a}{2} +0 \right) 
   +\psi' \left( \ds\frac{a}{2} -0 \right) \right\} =0 .
\end{eqnarray}
Here, $L^{(+)}_{1}$ and $L^{(-)}_{1}$ characterize the junction
conditions at $x=-a/2$, while $L^{(+)}_{2}$ and $L^{(-)}_{2}$ 
characterize those at $x=a/2$.
Solving Eqs.~(\ref{eq:jc-psi5})--(\ref{eq:jc-psi8}) 
under the assumption of Eq.~(\ref{eq:ef-psi2})
with respect to $A, B, C$ and $D$, we derive
\begin{eqnarray}
 \label{eq:dpi-A}
 A &= & \frac{e^{-ika}}{\Delta} \left\{ 
    -\left( 1+ikL_{2}^{(+)} \right) 
    \left( 1+ikL_{2}^{(-)} \right) \right.
    \nonumber \\
 & & \times 
    \left( 1+k^2 L_{1}^{(+)} L_{1}^{(-)} \right) 
    + \left( 1-ikL_{1}^{(+)} \right) 
    \left( 1-ikL_{1}^{(-)} \right) 
    \nonumber \\
 & & \left. \times
    \left( 1+k^2 L_{2}^{(+)} L_{2}^{(-)} \right) e^{2ika}
    \right\} , \\
 B & = & \frac{ik}{\Delta} 
    \left( L_{1}^{(+)} - L_{1}^{(-)} \right) 
    \left( 1+ikL_{2}^{(+)} \right) 
    \left( 1+ikL_{2}^{(-)} \right)  ,  \\
 C & = & -\frac{ik}{\Delta} 
    \left( L_{1}^{(+)} - L_{1}^{(-)} \right) 
    \left( 1+k^2 L_{2}^{(+)} L_{2}^{(-)} \right) 
    e^{ika}  , \\
 D & = & - \frac{k^2}{\Delta} 
    \left( L_{1}^{(+)} - L_{1}^{(-)} \right) 
    \left( L_{2}^{(+)} - L_{2}^{(-)} \right) ,
\end{eqnarray} 
where
\begin{eqnarray}
  \Delta & = & \left( 1+ikL_{1}^{(+)} \right) 
    \left( 1+ikL_{1}^{(-)} \right)
    \left( 1+ikL_{2}^{(+)} \right)  \nonumber \\
 &&  \times
    \left( 1+ikL_{2}^{(-)} \right)
    - \left( 1+k^2 L_{1}^{(+)} L_{1}^{(-)} \right) 
    \nonumber \\
 &&  \times
       \left( 1+k^2 L_{2}^{(+)} L_{2}^{(-)} \right) 
     e^{2ika}  .
\end{eqnarray}
Then, the transmission probability $T$ is calculated as
\begin{eqnarray}
 T_{2} = \left| D\right|^2 
 = \frac{ k^4  
    \left( L_{1}^{(+)} - L_{1}^{(-)} \right)^2 
    \left( L_{2}^{(+)} - L_{2}^{(-)} \right)^2}
    {\left| \Delta \right|^2} .
\end{eqnarray} 
If $L_{1}^{(+)} = L_{1}^{(-)}$ or 
$ L_{2}^{(+)} = L_{2}^{(-)}$, then the transmission 
probability completely vanishes in the same way 
as the case of a single point interaction.

\subsection{Conditions for resonant transmission}

\begin{figure}[tb]
  \includegraphics[width=.5 \textwidth]{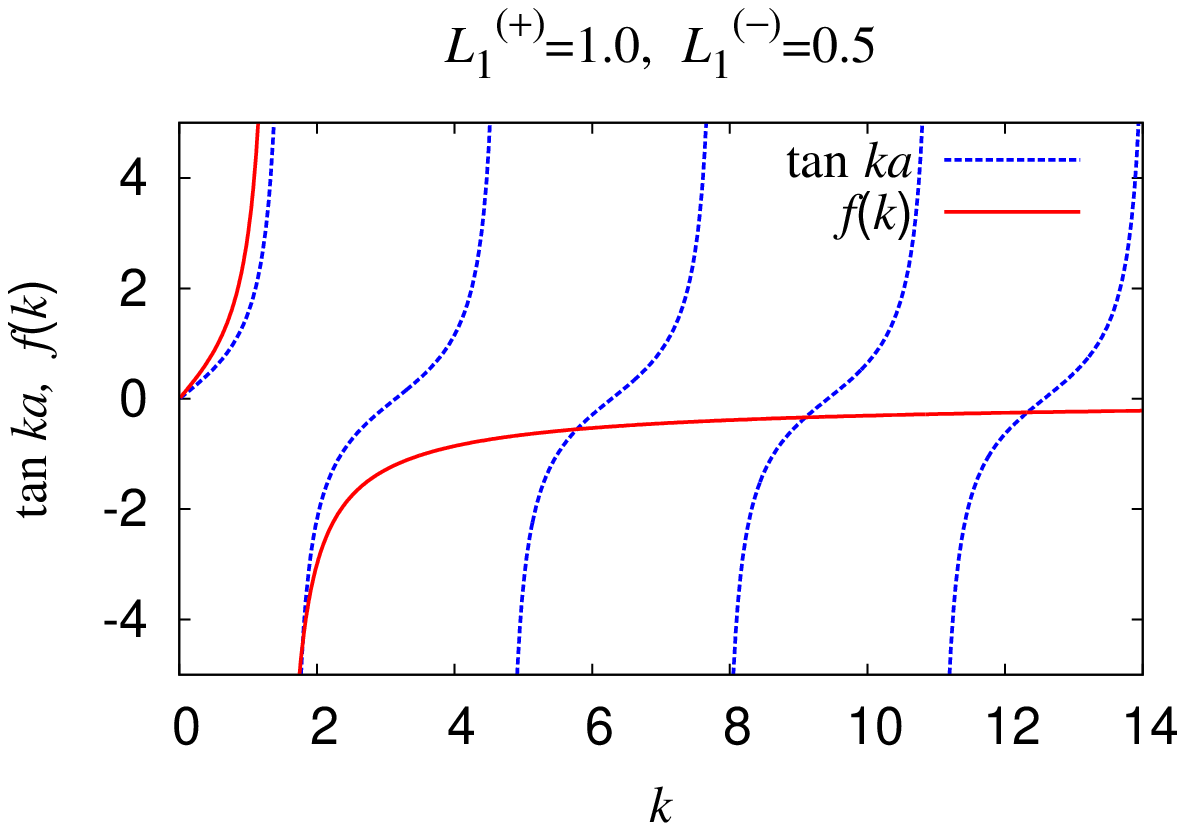}
  \caption{Functions in the resonance condition (\ref{eq:rc1})
    for $L_{1}^{(+)}+L_{1}^{(-)}>0$
    and $L_{1}^{(+)} L_{1}^{(-)}>0$.  
    Here, we adopt $a=1.0$, $L_{1}^{(+)}=1.0$ and $L_{1}^{(-)}=0.5$.
    The horizontal axis denotes the wave number $k$. 
    Perfect transmission occurs at the points of intersection 
    between the solid (red) curves and the dashed (blue) curves.}
  \label{fig3}
\end{figure}
\begin{figure}[tb]
  \includegraphics[width=.5 \textwidth]{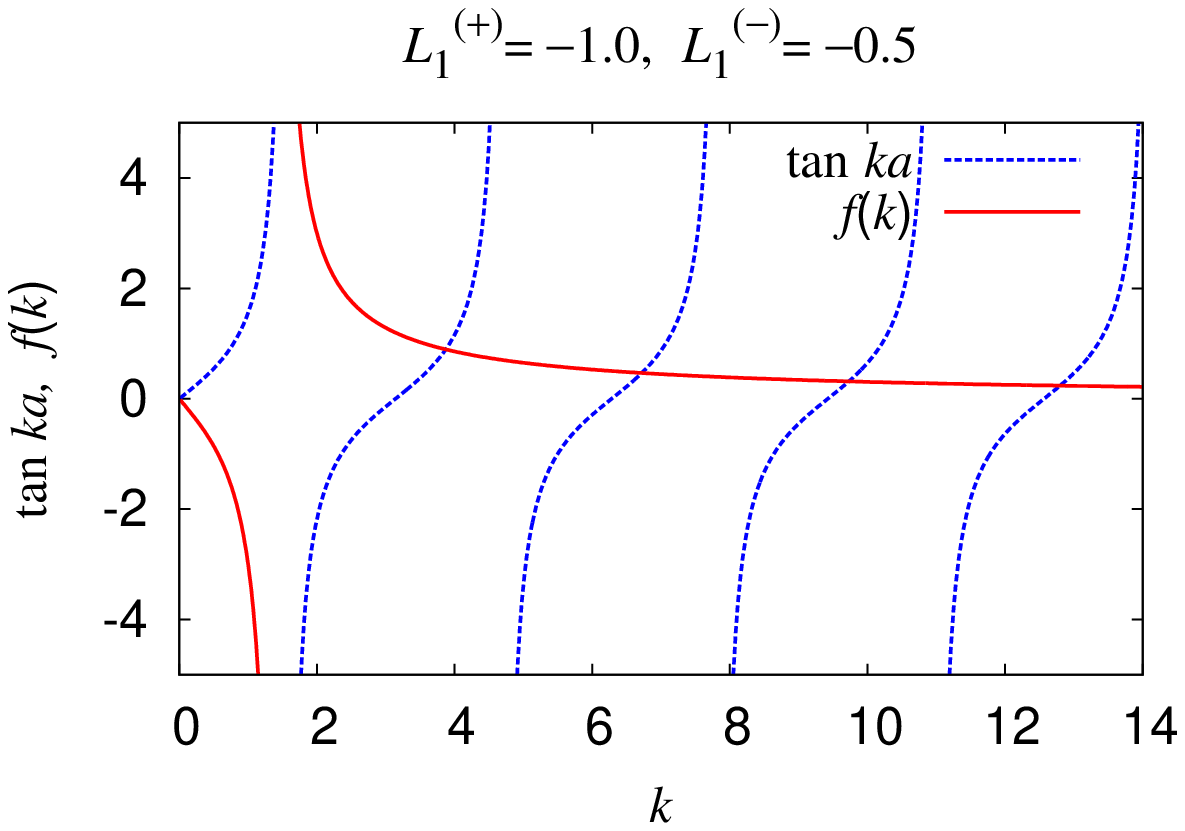}
  \caption{Functions in the resonance condition (\ref{eq:rc1})
    for $L_{1}^{(+)}+L_{1}^{(-)}<0$
    and $L_{1}^{(+)} L_{1}^{(-)}>0$.  
    Here, we adopt $a=1.0$, $L_{1}^{(+)}=-1.0$ and $L_{1}^{(-)}=-0.5$.
    The horizontal axis denotes the wave number $k$. 
    Perfect transmission occurs at the points of intersection 
    between the solid (red) curves and the dashed (blue) curves.}
  \label{fig4}
\end{figure}
\begin{figure}[htb]
  \includegraphics[width=.5 \textwidth]{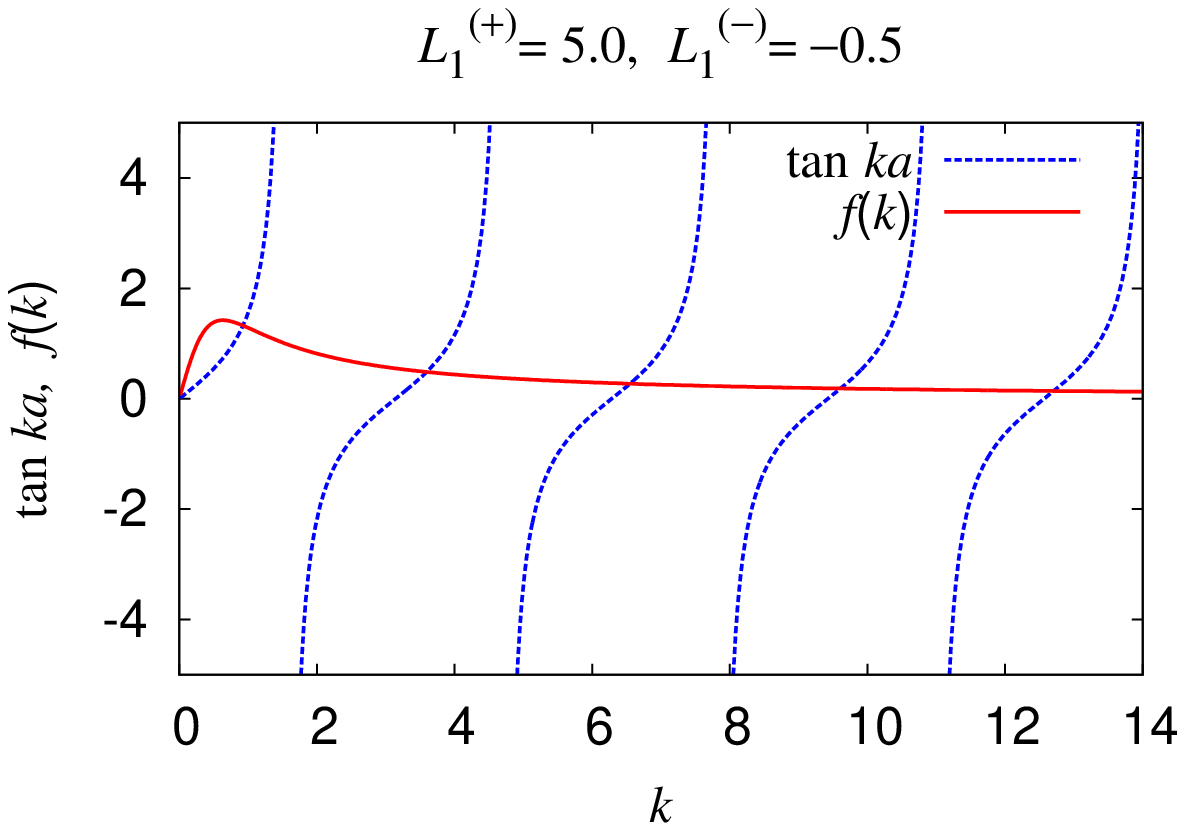}
  \caption{Functions in the resonance condition (\ref{eq:rc1})
    for $L_{1}^{(+)}+L_{1}^{(-)}>0$
    and $L_{1}^{(+)} L_{1}^{(-)}<0$.  
    Here, we adopt $a=1.0$, $L_{1}^{(+)}=5.0$ and $L_{1}^{(-)}=-0.5$.
    The horizontal axis denotes the wave number $k$. 
    Perfect transmission occurs at the points of intersection 
    between the solid (red) curves and the dashed (blue) curves.}
  \label{fig5}
\end{figure}
\begin{figure}[htb]
  \includegraphics[width=.5 \textwidth]{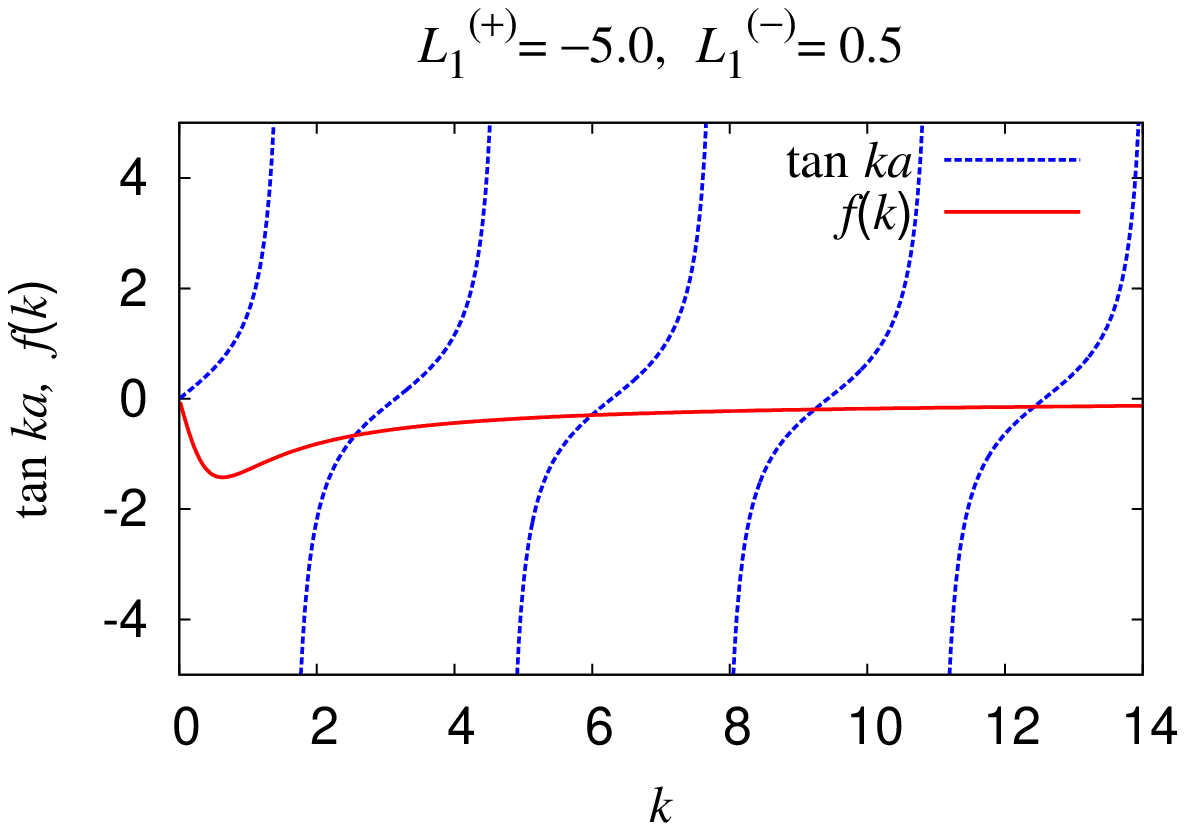}
  \caption{Functions in the resonance condition (\ref{eq:rc1})
    for $L_{1}^{(+)}+L_{1}^{(-)}<0$
    and $L_{1}^{(+)} L_{1}^{(-)}<0$.  
    Here, we adopt $a=1.0$, $L_{1}^{(+)}=-5.0$ and $L_{1}^{(-)}=0.5$.
    The horizontal axis denotes the wave number $k$. 
    Perfect transmission occurs at the points of intersection 
    between the solid (red) curves and the dashed (blue) curves.}
  \label{fig6}
\end{figure}

We investigate the conditions for perfect transmission.
 From the inequality $T_{2} \leq 1$, we obtain
\begin{eqnarray} 
 \lefteqn{\left(  M_{11} \sin ka  + M_{12} \cos ka \right)^2}
     \nonumber \\ 
 & + \left( M_{21} \sin ka
     + M_{22} \cos ka \right)^2 \geq 0 ,
\end{eqnarray}
where
\begin{eqnarray}
 M_{11} & = &  
  2 \left( 1- k^4 L_{1}^{(+)} L_{1}^{(-)} L_{2}^{(+)} L_{2}^{(-)}
  \right) , \\
 M_{12} & = &  -k L_{1}^{(+)} - k L_{1}^{(-)}
   - k L_{2}^{(+)}  - k L_{2}^{(-)}  
  \nonumber \\
  && 
   - k^3 L_{1}^{(+)} L_{1}^{(-)} L_{2}^{(+)} 
   - k^3 L_{1}^{(+)} L_{1}^{(-)} L_{2}^{(-)} 
   \nonumber \\
  && 
    - k^3 L_{1}^{(+)} L_{2}^{(+)} L_{2}^{(-)}
    - k^3 L_{1}^{(-)} L_{2}^{(+)} L_{2}^{(-)} ,
    \\
M_{21} & = & k L_{1}^{(+)} + k L_{1}^{(-)}
     - k L_{2}^{(+)} 
     - k L_{2}^{(-)} 
    \nonumber \\
    &&
    - k^3 L_{1}^{(+)} L_{1}^{(-)} L_{2}^{(+)} 
    - k^3 L_{1}^{(+)} L_{1}^{(-)} L_{2}^{(-)}
    \nonumber \\ 
    && 
     + k^3 L_{1}^{(+)} L_{2}^{(+)} L_{2}^{(-)}
     + k^3 L_{1}^{(-)} L_{2}^{(+)} L_{2}^{(-)}, \\
 M_{22} & = & 
   - 2 \left( k^2 L_{1}^{(+)} L_{1}^{(-)}  - k^2 L_{2}^{(+)} L_{2}^{(-)}
     \right) .
\end{eqnarray}
Thus, we derive the following conditions
for the perfect transmission, i.e., $T_{2}=1$,
\begin{eqnarray}
\label{eq:r-c1}
   M_{11} \sin ka + M_{12} \cos ka & = & 0 ,\\ 
\label{eq:r-c2}
   M_{21} \sin ka + M_{22} \cos ka & = & 0 .
\end{eqnarray}
These equations with respect to $k$ have solutions if and only if 
\begin{equation}
\label{eq:det-M}
 \det \left(
  \begin{array}{cc}
   M_{11} & M_{12} \\ M_{21} & M_{22}
  \end{array} 
 \right) =0 .
\end{equation}
Note that when this equation holds,
Eqs.~(\ref{eq:r-c1}) and (\ref{eq:r-c2}) 
give one independent equation.
The condition (\ref{eq:det-M}) is expressed as
\begin{equation}
\label{eq:det-M2}
 \alpha k^4 +2 \beta k^2 + \gamma =0 ,
\end{equation}
where
\begin{eqnarray}
 \alpha & = & 
  \left( L_{1}^{(+)} - L_{1}^{(-)} \right)^2
  \left( L_{2}^{(+)} L_{2}^{(-)}\right)^2 
  \nonumber \\
 && 
  - \left( L_{2}^{(+)} - L_{2}^{(-)} \right)^2
  \left( L_{1}^{(+)} L_{1}^{(-)}\right)^2 , \\
 \beta & = &
  \left( L_{1}^{(+)} - L_{1}^{(-)} \right)^2
  L_{2}^{(+)} L_{2}^{(-)} 
  - \left( L_{2}^{(+)} - L_{2}^{(-)} \right)^2
  \nonumber \\
 && \times
  L_{1}^{(+)} L_{1}^{(-)} , \\
 \gamma & = &
  \left( L_{1}^{(+)} - L_{1}^{(-)} \right)^2
  - \left( L_{2}^{(+)} - L_{2}^{(-)} \right)^2 .
\end{eqnarray}
When all of the coefficients in Eq.~(\ref{eq:det-M2}) 
vanish, i.e.,
\begin{equation}
 \label{eq:inf-s-c}
 \alpha = \beta = \gamma =0,  
\end{equation}
Eq.~(\ref{eq:det-M}) is identically satisfied, 
independent of the value of $k$.
Equation (\ref{eq:inf-s-c}) gives
\begin{equation}
 L_{1}^{(+)} L_{1}^{(-)} 
  = L_{2}^{(+)} L_{2}^{(-)} ,
\end{equation}
\begin{equation}
 \left( L_{1}^{(+)} \right)^2
 + \left( L_{1}^{(-)} \right)^2
 = \left( L_{2}^{(+)} \right)^2
 + \left( L_{2}^{(-)} \right)^2 ,
\end{equation}
which leads to the relations
\begin{equation}
\label{eq:inf-s-c2}
 \left( 
  \begin{array}{c}
   L_{2}^{(+)} \\  L_{2}^{(-)}
  \end{array}
 \right) = \pm
 \left( 
   \begin{array}{c}
    L_{1}^{(+)} \\ L_{1}^{(-)} 
   \end{array}
 \right), \quad \mbox{or} \quad 
 \left( 
  \begin{array}{c}
   L_{2}^{(+)} \\  L_{2}^{(-)}
  \end{array}
 \right) = 
 \pm \left( 
   \begin{array}{c}
    L_{1}^{(-)} \\ L_{1}^{(+)} 
   \end{array}
 \right) .
\end{equation}
Therefore, when the relations (\ref{eq:inf-s-c2}) hold, 
the necessary and sufficient condition (\ref{eq:det-M}) 
is identically satisfied. Then, we can generally obtain 
solutions for the perfect transmission 
by solving Eq.~(\ref{eq:r-c1}) or (\ref{eq:r-c2}).

We investigate all the cases in Eq.~(\ref{eq:inf-s-c2})
in the following.

(i) The cases of 
$\ds \left( L_{2}^{(+)}, L_{2}^{(-)} \right)
   = \left( L_{1}^{(+)}, L_{1}^{(-)} \right)$ or
$\left( L_{1}^{(-)}, L_{1}^{(+)} \right)$.
From Eq.~(\ref{eq:r-c1}) or (\ref{eq:r-c2}), we derive
\begin{eqnarray}
\label{eq:rc0}
 \lefteqn{\left( 1 + k^2  L_{1}^{(+)} L_{1}^{(-)} \right)
  \left\{ \left( 1 - k^2  L_{1}^{(+)} L_{1}^{(-)} \right) 
  \sin ka \right.}
  \nonumber \\
 & \left.
  - k \left( L_{1}^{(+)} + L_{1}^{(-)} \right) \cos ka \right\}
  = 0 .
\end{eqnarray}
If $L_{1}^{(+)} L_{1}^{(-)} < 0$, then we find 
a solution 
\begin{equation}
\label{eq:pt}
 k= \sqrt{-\frac{1}{L_{1}^{(+)} L_{1}^{(-)}}}.
\end{equation}
This result is the same as 
in the case of a single point interaction.
We can also find an infinite number of solutions
for perfect transmission through the condition
derived from Eq.~(\ref{eq:rc0}), 
\begin{equation}
 \label{eq:rc1}
 \tan ka = f(k) ,
\end{equation}
where
\begin{equation}
 f(k):=\frac{k \left( L_{1}^{(+)} + L_{1}^{(-)} \right)}
   {1-k^2 L_{1}^{(+)} L_{1}^{(-)}} .
\end{equation}
The behavior of the function $f(k)$ depends on
the signs of $L_{1}^{(+)} + L_{1}^{(-)}$ and
$L_{1}^{(+)} L_{1}^{(-)}$. Representative examples in
each cases are shown in Figs.~\ref{fig3}--\ref{fig6}, 
In these figures, we plot the curves of 
the functions on the both sides in Eq.~(\ref{eq:rc1}).
At the points of intersection between the solid (red) curves 
and the dashed (blue) curves, perfect transmission occurs.
Consequently, we can find an infinite number of solutions
for perfect transmission.

(ii) The cases of 
$\ds \left( L_{2}^{(+)}, L_{2}^{(-)} \right)
   = \left( -L_{1}^{(+)}, -L_{1}^{(-)} \right)$ or
$\left( -L_{1}^{(-)}, -L_{1}^{(+)} \right)$.
From Eq.~(\ref{eq:r-c1}) and (\ref{eq:r-c2}), we have
\begin{eqnarray}
 \label{eq:pt2-1}
 \left( 1 - k^2  L_{1}^{(+)} L_{1}^{(-)} \right)  
  \left( 1 + k^2  L_{1}^{(+)} L_{1}^{(-)} \right) 
  \sin ka = 0 .
\end{eqnarray}
\begin{eqnarray}
 \label{eq:pt2-2}
 k \left( L_{1}^{(+)} + L_{1}^{(-)} \right) 
  \left( 1 + k^2  L_{1}^{(+)} L_{1}^{(-)} \right) 
  \sin ka = 0 .
\end{eqnarray}
If $L_{1}^{(+)} + L_{1}^{(-)}=0$, 
we have $L_{1}^{(+)} L_{1}^{(-)}<0$
and $\left( 1 - k^2  L_{1}^{(+)} L_{1}^{(-)} \right) >0$.
Thus, from Eq.~(\ref{eq:pt2-1}), we derive 
\begin{eqnarray}
\label{eq:pt2-3}
  \left( 1 + k^2  L_{1}^{(+)} L_{1}^{(-)} \right) 
  \sin ka = 0 .
\end{eqnarray}
If $L_{1}^{(+)} + L_{1}^{(-)} \neq 0$, then we derive 
Eq.~(\ref{eq:pt2-3}) again from Eq.~(\ref{eq:pt2-2}).
It follows that if $L_{1}^{(+)} L_{1}^{(-)} < 0$, 
we find the solution (\ref{eq:pt}) again.
We also find an infinite number of solutions from the condition
\begin{equation}
 \sin ka = 0 .
\end{equation}
This leads to the solutions 
\begin{equation}
 k = \frac{n \pi}{a} \quad \left( n=1,2,3,\cdots \right) .
\end{equation}
for perfect transmission.

\begin{figure}[htb]
  \includegraphics[width=.5 \textwidth]{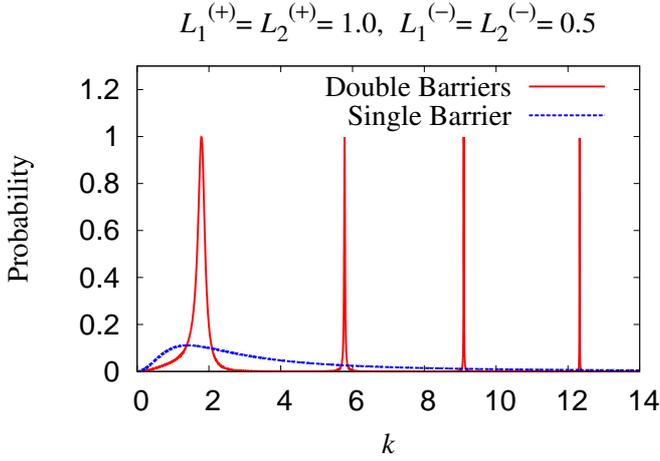}
  \caption{The transmission probability for double barriers
    is shown as a function of $k$ by the solid (red) curve, 
    when $L_{1}^{(+)}=L_{2}^{(+)}=1.0$ and 
    $L_{1}^{(-)}=L_{2}^{(-)}=0.5$.
    Here, we adopt $a=1.0$. The perfect transmission ($T_{2}=1$)
    occurs when the condition $\tan ka =f(k)$ is satisfied.
    The transmission probability for the single barrier 
    with $L_{1}^{(+)}$ and $L_{1}^{(-)}$ is also shown by 
    the dashed (blue) curve.}
  \label{fig7}
\end{figure}
\begin{figure}[htb]
  \includegraphics[width=.5 \textwidth]{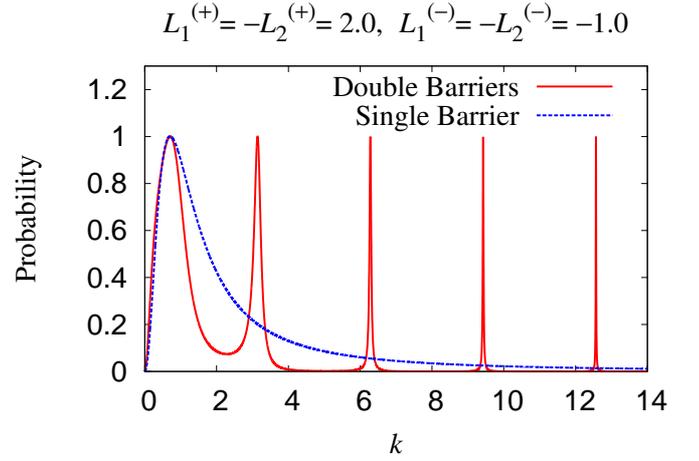}
  \caption{The transmission probability for double barriers
    is shown as a function of $k$ by the solid (red) curve, 
    when $L_{1}^{(+)}=-L_{2}^{(+)}=2.0$ and 
    $L_{1}^{(-)}=-L_{2}^{(-)}=-1.0$.
    Here, we adopt $a=1.0$. The first peak appears at 
    $k=\sqrt{-1/(L_{1}^{(+)} L_{1}^{(-)})}$, while the other
    peaks appear at $k=n\pi / a$.
    The transmission probability for the single barrier 
    with $L_{1}^{(+)}$ and $L_{1}^{(-)}$ is also shown by 
    the dashed (blue) curve.}
  \label{fig8}
\end{figure}

We show representative examples of the transmission probability
as a function of $k$ for the above cases
in Figs.~\ref{fig7} and \ref{fig8}.
In these figures, we show the transmission probability 
for double barriers by the solid (red) curves.
We also show the transmission probability for a single barrier
by the dashed (blue) curves for comparison.
In Fig.~\ref{fig7}, we adopt $a=1.0$, $L_{1}^{(+)}=L_{2}^{(+)}=1.0$ 
and $L_{1}^{(-)}=L_{2}^{(-)}=0.5$, while in Fig.~\ref{fig8}, 
we adopt $a=1.0$, $L_{1}^{(+)}=-L_{2}^{(+)}=2.0$ 
and $L_{1}^{(-)}=-L_{2}^{(-)}=-1.0$. 
In these cases, we can confirm the periodic resonant peaks, 
at which perfect transmission occurs.
Furthermore, we find that the peak width decreases as $k$
increases. In particular, when $L_{1}^{(+)}=-L_{2}^{(+)}$ 
and $L_{1}^{(-)}=-L_{2}^{(-)}$, the transmission probability
$T_{2}$ can be expanded around a peak as 
\begin{equation}
 T_{2}(k) \simeq 1- \left( \frac{k-k_{n}}{w} \right)^2 + \cdots ,
\end{equation}
where $k_{n} = n\pi/a$ and 
\begin{equation}
 w=\frac{k_{n} \left( L_{1}^{(+)}- L_{1}^{(-)} \right)}
   {2a \left( 1+k_{n}^2 L_{1}^{(+)} L_{1}^{(-)} \right)}
   \sqrt{T_{1}\left(k_{n},L_{1}^{(+)}, L_{1}^{(-)} \right)}
\end{equation}
Here, $T_{1}$ is given by Eq.~(\ref{eq:t-single}).
The peak width is roughly given by $w$.
Therefore, the peak width is proportional to the 
square root of the transmission probability 
for a single barrier and decreases as $k_n$ increases.
Similar feature could be found also in the case of 
$L_{1}^{(+)}=L_{2}^{(+)}$ and $L_{1}^{(-)}=L_{2}^{(-)}$.

Let us reconsider the results of Eq.~(\ref{eq:inf-s-c2})
concretely from the view point of potential functions.
For example, when we assume a delta function potential
at $x=-\frac{a}{2}$, i.e., 
$V(x)=-\frac{\hbar^2}{mL_{1}^{(+)}} \delta (x+\frac{a}{2})$,
which is given by $L_{1}^{(+)} \neq 0$
and $L_{1}^{(-)}= 0$, an infinite number of resonant peaks
appear in the following four cases:
\begin{enumerate}
 \item[(I)] $\left( L_{2}^{(+)} , L_{2}^{(-)}\right) 
  = \left( L_{1}^{(+)} , 0\right)$, 
 \item[(II)] $\left( L_{2}^{(+)} , L_{2}^{(-)}\right) 
  = \left( - L_{1}^{(+)} , 0\right)$, 
 \item[(III)] $\left( L_{2}^{(+)} , L_{2}^{(-)}\right) 
  = \left( 0, L_{1}^{(+)} \right)$, 
 \item[(IV)] $\left( L_{2}^{(+)} , L_{2}^{(-)}\right) 
  = \left( 0, - L_{1}^{(+)} \right)$.
\end{enumerate}
The cases (I) and (II) correspond to
the potentials $-\frac{\hbar^2}{mL_{1}^{(+)}} \delta (x-\frac{a}{2})$
and $\frac{\hbar^2}{mL_{1}^{(+)}} \delta (x-\frac{a}{2})$,
respectively. These might be predictable consequences.
However, the last two cases (III) and (IV) would be 
unexpected results.

Finally, it should be noticed that even if Eq.~(\ref{eq:inf-s-c}) 
does not hold, the positive solution $k$ satisfying 
the condition (\ref{eq:det-M}) or (\ref{eq:det-M2})
may exist when the solution of Eq.~(\ref{eq:det-M2})
\begin{eqnarray}
  k^2 = \frac{-\left( L_{2}^{(+)} - L_{2}^{(-)} \right)
   \pm \left( L_{1}^{(+)} - L_{1}^{(-)} \right)}
{L_{1}^{(+)} L_{1}^{(-)} \left( L_{2}^{(+)} - L_{2}^{(-)} \right)
 - L_{2}^{(+)} L_{2}^{(-)} \left( L_{1}^{(+)} - L_{1}^{(-)} \right)}
\end{eqnarray}
is positive. In this case, Eq.~(\ref{eq:r-c1}) and 
(\ref{eq:r-c2}) could be satisfied 
for a specific value of $a$. Then,  
the perfect transmission would occur incidentally
in this case.

\section{Concluding remarks}
\label{sec4}

We have considered the scattering of a quantum particle
by two independent, successive parity-invariant point interactions
in one dimension. The parameter space is given by the direct
product of two tori and described by four parameters
$L_{1}^{(+)}$, $L_{1}^{(-)}$, $L_{2}^{(+)}$ and $L_{2}^{(-)}$.
By considering incident plane wave, 
we derived the formula for the transmission probability 
without any assumptions about the parameter space.
Based on the formula, we investigated the conditions
for the parameter space under which the perfect resonant 
transmission occur. Finally, we found the resonance
conditions, which are the main results in this paper, 
to be given by the symmetric and anti-symmetric relations 
(\ref{eq:inf-s-c2}) between the parameters

In this paper, we restricted our attention
to the parity-invariant point interactions.
When we relax this assumption, the parameter space
becomes larger, i.e.,  $U(2) \otimes U(2)$.
This extension will be discussed elsewhere \cite{knt2}.
Furthermore, the properties of resonant transmission 
through $N$ independent multiple point interactions 
would be future works.

%%%6
Finally, it should be noted that the analysis of 
our physical systems from the viewpoint of the $S$ matrix 
on the complex $k$-plane would also be important future works.
From this approach, we could discuss quasi-stationary 
or resonance states which appear between the two 
potential barriers, and its lifetime.  
The authors of \cite{hernandez,antoniou} investigated 
the poles of $S$ matrix in the system of a double delta 
barrier potential. Our physical systems in the present
paper give the extension of their system.
Therefore, the analysis based on the $S$ matrix 
would give us a deep understanding of the physical processes. 
%%%

%% The Appendices part is started with the command \appendix;
%% appendix sections are then done as normal sections
%% \appendix

%% \section{}
%% \label{}

%% If you have bibdatabase file and want bibtex to generate the
%% bibitems, please use
%%
%%  \bibliographystyle{elsarticle-num} 
%%  \bibliography{<your bibdatabase>}

%% else use the following coding to input the bibitems directly in the
%% TeX file.

\end{document}